\documentclass[sigconf]{acmart}
\AtBeginDocument{%
  \providecommand\BibTeX{{%
    \normalfont B\kern-0.5em{\scshape i\kern-0.25em b}\kern-0.8em\TeX}}}

\copyrightyear{2022}
\acmYear{2022}
\setcopyright{acmcopyright}
\acmConference[WiseML '22] {Proceedings of the 2022 ACM Workshop on Wireless Security and Machine Learning}{May 19, 2022}{San Antonio, TX, USA.}
\acmBooktitle{Proceedings of the 2022 ACM Workshop on Wireless Security and Machine Learning (WiseML '22), May 19, 2022, San Antonio, TX, USA}
\acmPrice{15.00}
\acmISBN{978-1-4503-9277-8/22/05}
\acmDOI{10.1145/3522783.3529519}

\begin{document}

\fancyhead{}
\title{Digital Twin Virtualization with Machine Learning for IoT and Beyond 5G Networks: Research Directions for Security and Optimal Control}


\author{Jithin Jagannath\textsuperscript{1}, Keyvan Ramezanpour\textsuperscript{1}, Anu Jagannath\textsuperscript{1}   }
\affiliation{%
  \institution{
  \textsuperscript{1}Marconi-Rosenblatt AI/ML Innovation Lab, ANDRO Computational Solutions LLC} 
  \country{Rome, NY, USA}
}
\email{{jjagannath, kramezanpour, ajagannath}@androcs.com}

\renewcommand{\shortauthors}{Jagannath and Ramezanpour, et al.}
\begin{abstract}
  Digital twin (DT) technologies have emerged as a solution for real-time data-driven modeling of cyber physical systems (CPS) using the vast amount of data available by Internet of Things (IoT) networks. In this position paper, we elucidate unique characteristics and capabilities of a DT framework that enables realization of such promises as online learning of a physical environment, real-time monitoring of assets, Monte Carlo heuristic search for predictive prevention, on-policy, and off-policy reinforcement learning in real-time. We establish a conceptual layered architecture for a DT framework with decentralized implementation on cloud computing and enabled by artificial intelligence (AI) services for modeling and decision-making processes. The DT framework separates the control functions, deployed as a system of logically centralized process, from the physical devices under control, much like software-defined networking (SDN) in fifth generation (5G) wireless networks. To clarify the significance of DT in lowering the risk of development and deployment of innovative technologies on existing system, we discuss the application of implementing zero trust architecture (ZTA) as a necessary security framework in future data-driven communication networks.
\end{abstract}




\maketitle

\section{Introduction}
Evolution of IoT networks has been a key enabler of modern technological developments including smart grid and city infrastructure, intelligent transportation, smart healthcare, autonomous industrial plants, surveillance and intelligent command and control in the military section, to name a few. Massive number of IoT devices, potentially distributed over a large-scale physical environment, lead to a vast amount of data. Hence, intelligent technologies employing IoT networks face the major challenges of communication, processing and interpretation of big data. 

The fifth-generation (5G) mobile networks provide IoT devices with low latency and seamless connectivity for handling big data. Artificial intelligence (AI) and machine learning (ML) engines realize efficient algorithms for processing the data \cite{jagannath2019machine,jagannath2021redefining}. Digital twin (DT) technology addresses the interpretation of \textit{distributed data} within a certain context. It enables extracting relevant information about the environment required by the upper-layer decision-making and control processes. The DT technology thus enables the realization of the so-called hyper-automation \cite{jacoby2020digital}.

Recent research trend suggests that DT technology is becoming a critical part of any large-scale intelligent system that requires real-time monitoring and interaction with the environment.
The role of DT technology in distributed intelligent systems can be compared with the software defined networking (SDN) architecture in 5G networks. In a network architecture based on SDN, the control plane (traffic management) is separated from data plane (traffic flow) and realized in a logically centralized control process \cite{yousaf2017nfv}. Hence, the SDN allows realization of a programmable network connectivity with dynamic management of user traffic demands. With implementation on cloud computing platforms, the SDN has led to realization of carrier clouds in cellular networks which allows development of \textit{softwarized} services over existing infrastructure and hardware equipment. 

Similar to SDN, the DT technology enables realization of optimal decision-making and control processes in a logically centralized process, possibly deployed on cloud platforms. It separates the control process (in the \textit{virtual space}) from the data plane defining the flow of measurements collected from the environment by IoT devices and their interaction with the environment (in the \textit{physical space}). As a result, DT technology generalizes network connectivity of SDN and realizes a wide range of programmable functions on the same physical IoT infrastructure.

Another aspect of DT technology is \textit{virtualization} of physical objects and processes. This view is in parallel with network function virtualization (NFV) in 5G networks. The NFV refers to the realization of network functions, including evolved packet core functions, routers, and firewalls, in software for execution on virtual machines rather than specialized hardware equipment \cite{han2015network}. Hence, it significantly improves the flexibility and agility in development and deployment of network core on general purpose servers and cloud infrastructure. The NFV thus addresses the scalability issue of 5G networks in providing a massive volume of mobile and IoT devices with connectivity. DT models extend the virtualization of logic functions, as in 5G core, to any physical object or process. Hence, DT technologies provide a scalable solution for real-time monitoring and deployment of event detection and control processes for a massive volume of IoT devices.

An interesting perspective on DT technologies can be obtained by studying their capability in realizing the concepts of \textit{learning} and \textit{planning} in the context of artificial intelligence. Learning refers to improving a control policy toward the optimal solution using direct observations and interactions with a system or environment. Planning refers to finding the optimal control policy using simulated interactions with the environment according to a distribution or sample model of the environment. The cycle of direct learning, model construction and planning can significantly accelerate convergence to the optimal policy. A DT framework is a realization of the model construction and simulation steps. Hence, a DT framework supports realization of a unified intelligent control process involving both direct learning and planning.

In this paper, we introduce an intelligent DT framework as a scalable solution for data-driven modeling and live simulation of large-scale systems over 5G/6G based IoT networks. The DT framework adopts a layered architecture for distributed deployment on cloud computing platforms. It also enables the implementation of AI/ML engines for event detection, predictive models, learning, and planning for a wide range of services and functions within a unified framework. We also study development and deployment of intelligent zero trust architectures (ZTA) as a use case of the proposed DT in enabling rapid prototyping, testing, risk assessment, and integration of innovative security solutions into existing large-scale systems. 

The rest of the paper is organized as follows. A review of definitions, role, and emerging applications of DT is presented in Section \ref{sec:role}. The envisioned architecture of DT for cloud-based deployment is described in Section \ref{sec:arch}. Research directions for network security is discussed in Section \ref{sec:directions} and the paper concludes in Section \ref{sec:conclusions}.

\section{Definition and Role of Digital Twin} \label{sec:role}

There are multitude of different, but similar, definitions for DT in the literature, reflecting adoption of DT in various research domains. A common definition of DT capturing most of the concepts used in various applications is “a probabilistic simulation model for an as-built physical entity (sensor, device, system or process) that represents the properties (control and configuration capabilities), state (updates), conditions (interactions with the environment), and history of the entity in a way that mirrors its life and behavior”. A comprehensive list of definitions in different applications is given in \cite{liu2022state}. In this paper, rather than the exact definition we elucidate the main characteristics and requirements of DTs.

A DT is a unified high-fidelity simulation model that represents a physical entity in the past, present, and future. With respect to the past and the present, a DT in the virtual space mirrors the behavior of an entity in the physical space. Regarding the future, the DT predicts accurately the behavior of the entity which is sufficient for the control process. Further, the behavior of the entity in a simulated environment (which is not necessarily observed) can be inferred using the DT models for the purpose of policy exploration. 

\begin{figure}[t!]
	\centering
		\includegraphics[width=0.45\textwidth]{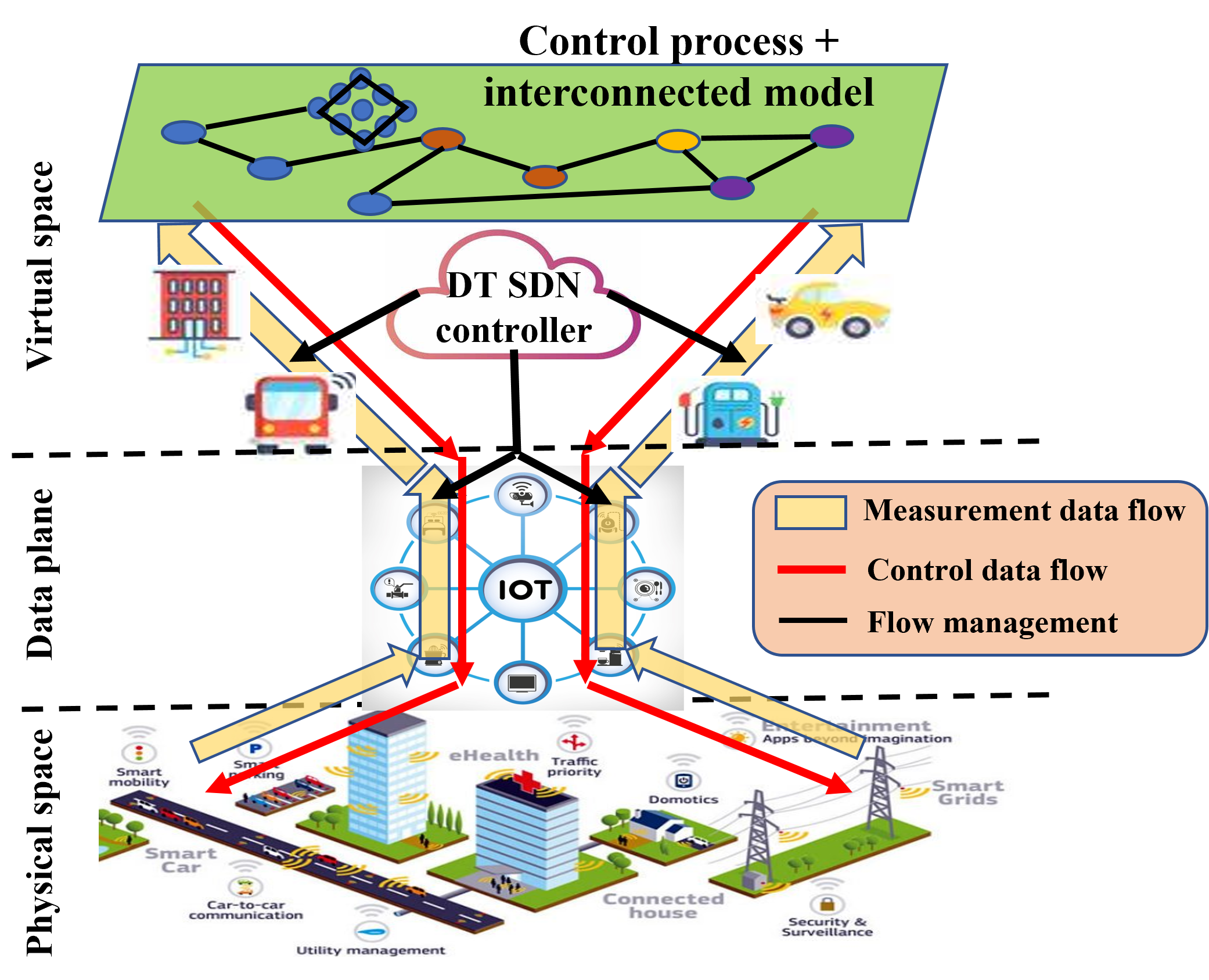}
	\caption{Overall concept of digital twin (DT) in separating control process (in virtual space) from physical devices through IoT network (data plane). }
	\label{fig:concept}
\end{figure}

Constructing a model with the above capabilities requires a bidirectional communication between the physical entity and the DT. The measurements are collected from the physical space, transferred through the network, and consumed by DTs in real-time to update models, and potentially control policies. Control commands computed in the virtual space of DTs are also sent to the physical space through the network. This relationship is depicted in Fig. \ref{fig:concept} in which the data plane represents the networking mechanism (IoT network). In fact, real-time data communication and online learning of models are the distinctive features of a DT framework compared with traditional static simulation frameworks. The latter includes models of physical entities in isolation from the environment while a DT framework additionally learns the environment and updates the models accordingly.

\subsection{Key Capabilities of DT} \label{sec:capabilities}

\begin{figure}[t!]
	\centering
		\includegraphics[width=0.3\textwidth]{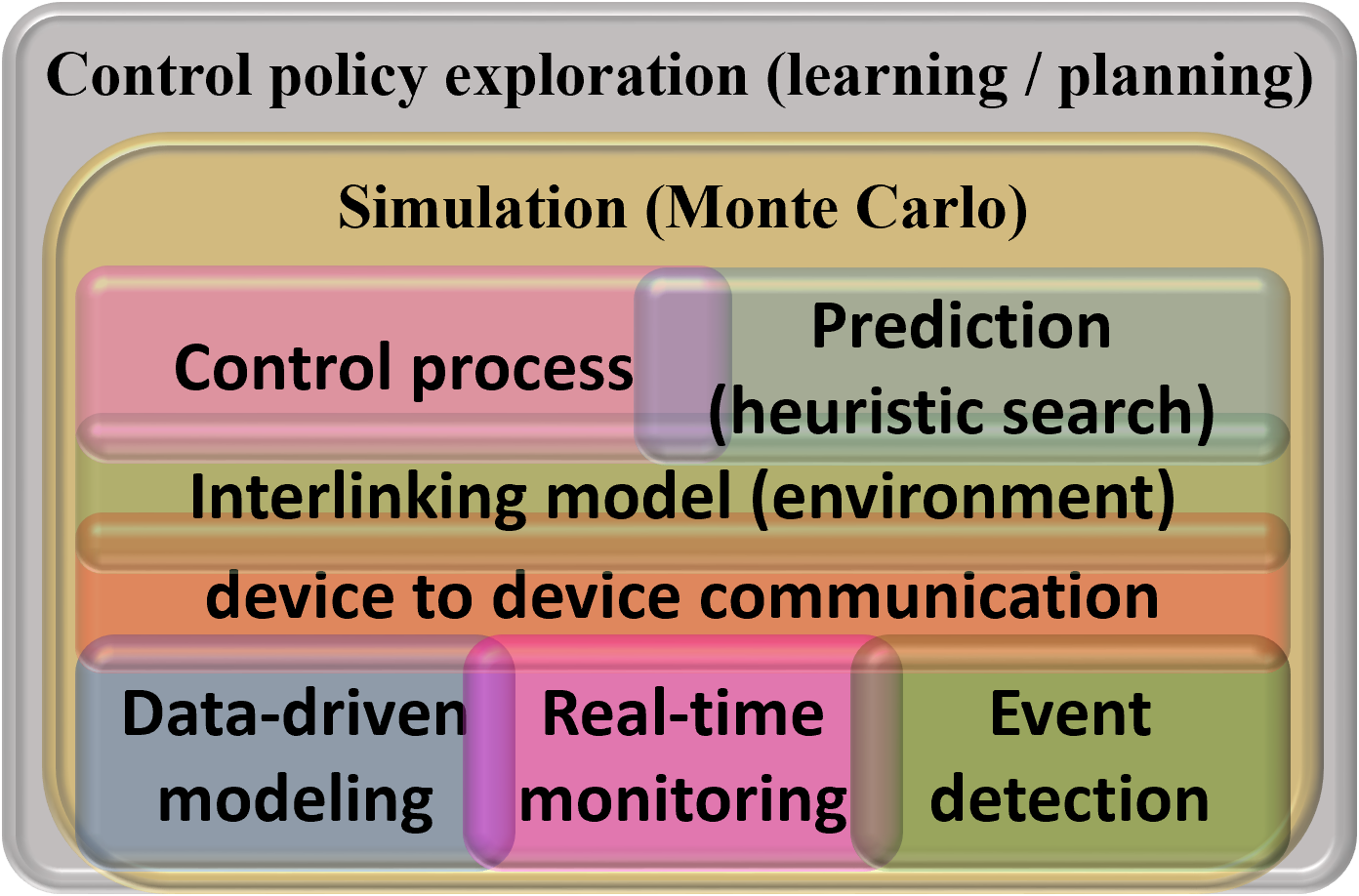}
	\caption{Characteristics and capabilities of DT in real-time monitoring, live simulation, learning and planning. }
	\label{fig:features}
\end{figure}

A DT framework provides a high-level representation of a physical space useful for the control process. The key capabilities expected of a DT framework are described below and summarized in Fig.~\ref{fig:features}. While these are not a comprehensive list of DT functionalities, they reveal the main role and motivation for the adoption of DT in various applications.

The classical role of DT is \textit{real-time monitoring} of remote and distributed devices over an IoT network. It provides users with the current status and historic conditions and a data description for machine-readable application programming interface (API). Another important role  of DT is realizing a logic function for \textit{event detection}. The logic can be a simple threshold detection when the state of the environment (or device) is described by a scalar variable. Deep neural networks are used in more advanced DT models for environments with a high-dimensional state space. 

In addition to above roles, an initial motivation for development of DT was to provide \textit{interoperability}. This property enabled defining a unified data description for heterogeneous IoT devices with a wide range of APIs and data formats. Hence, it could facilitate device to device communication. A DT framework also provides an \textit{interlinking model} that describes relationships between different types of data collected from many IoT devices distributed over an environment. The interlinking model presents a high-level understanding of the environment dynamics that is consumed by the control process. Modern DT frameworks are also expected to provide a \textit{simulation} environment that models the dynamics of the entire physical space sufficient for learning the optimal control policy for a given objective. Hence, a DT framework enables realization of online learning techniques based on planning such as Monte Carlo tree search (MCTS).

\subsection{Role of DT Technology}
Real-time monitoring and diagnostics have been a major propellant of DT technology in cyber-physical systems (CPS) and industrial IoT (IIoT) \cite{he2018surveillance}. The application of DT in medical CPS has been proposed as a framework for data collection and communication with cloud services, for processing and storage, over IoT networks \cite{jimenez2020health, alam2017c2ps}. Similarly, IIoT systems employ DT to monitor the health of machinery and detect faults in automated production plants \cite{tao2018digital}. In more advanced applications, DT has been employed in smart cities and smart grid systems to model and monitor resource and energy usage patterns \cite{baboli2020measurement, zhou2019digital,brosinsky2018recent}.

A popular application of DT is in the field of cybersecurity for anomaly detection, intrusion detection and prevention systems (IDS/IPS). In \cite{danilczyk2021smart}, a logic based on deep neural networks uses the DT models to detect anomaly in the smart grid infrastructure. Similarly, \cite{saad2020implementation} employs a cloud-based DT as a distributed framework for detecting and mitigating individual and coordinated security attacks on smart grid infrastructure. In these applications, DT serves as a model for a large-scale and distributed system that can be used by upper-layer detection and decision-making processes \cite{jiang2021novel}.

Evolution of DT technology with integration of AI/ML algorithms has led to the capability of predictive analytics or prognostics beyond the real-time event detection and diagnostics \cite{kaur2020convergence}. DT is used in \cite{liu2019data} to provide a fault prediction model, for the example of aero-engine bearing. The simulation functions of the DT are then used to explore the optimal maintenance strategy for identified faults. Similar concepts and methodology are introduced in \cite{damjanovic2018digital} to predict and mitigate the sources of privacy leakage in smart automotive systems. A DT framework employing a data-driven ML model of the smart grid and model-based state estimator has been proposed in \cite{tzanis2020hybrid} for fault prediction in near real-time and taking preventive actuation.

Interoperability is the key requirement of DT that allows realization of hyper-automation and intelligence in large-scale distributed systems. It has also been a major motivation of DT adoption in IoT/IIoT. Following this perspective, \cite{bellavista2021application} employs a DT framework as a networking middleware to facilitate interaction between heterogeneous IoT devices using IP-based protocols of the framework for communication between DT models. The DT also supports deployment of different applications (or services) on the same IoT platform by controlling communication mechanisms suitable for the target application.

A DT framework can implement multiple services on the same physical IoT infrastructure. This can be compared with \textit{network slicing} in the architecture of 5G and beyond networks. As an example, the IoT network in a smart grid can be used for learning energy consumption patterns and demands, modelling electric equipment, and detecting cybersecurity attacks as different services. These services constitute different slices of the DT framework controlling the smart grid. Deployment, upgrade or update of the services do not require reconfiguration of the IoT network nor the physical system. An interlinking model of DT provides a description of the environment sufficient for the objective of the control process within a target service.

Real-time learning and decision-making have been the critical component of modern large-scale and dynamic systems. A prominent example of intelligence in standardized systems is the radio access network (RAN) intelligent controller (RIC) in 5G networks. The RIC is the key component of the Open RAN (O-RAN) architecture for 5G network optimization. The RIC, and specifically AI engines of the RAN, optimizes critical RAN functions such as resource allocation, traffic prediction, mobility management, improving energy-efficiency and overall QoS and QoE \cite{balasubramanian2021ric}. The simulation environment of DT provides a platform for realizing policy exploration for intelligent control processes on distributed systems.

The realization of intelligent networking through DT is studied in \cite{zhao2020intelligent} for vehicular ad hoc networks (VANET). The (logically) centralized software-defined controller of the VANET implements learning-based networking schemes in response to the dynamics of the vehicles and network environment. These schemes include adaptive load balancing and scheduling, routing policy exploration, and flow-table construction. The DT framework provides a network model for predictive verification of routing policies. More importantly, the DT is used as a simulation platform (virtual space) for reinforcement learning that explores and learns the optimal routing policy before deployment on the real physical network.

Resource allocation in IoT networks, supporting a massive volume of heterogeneous devices (with diverse communication and computational capabilities), is a challenging task that requires adaptive and intelligent resource allocation. The dynamic nature of IoT networks, with ever increasing number of devices sharing limited network resources, has pushed online learning methods to model the network environment and explore resource management policies. In this regard, \cite{lu2020communication} employs a digital twin of the edge network for simulating bandwidth resource allocation and user scheduling. A reinforcement learning agent uses the simulation environment for policy learning. Further, a federated learning scheme has been employed for communication efficiency which also improves data security and privacy by avoiding sharing of raw data. 

DT frameworks have also been employed to model the IoT networks (data plane in Fig. \ref{fig:concept}) used for data exchange.
A DT model of IoT networks is employed in \cite{dai2020deep} for the purpose of computation offloading. A reinforcement learning agent uses the DT models of the network to distribute computation demands of devices for efficient use of IoT computing resources. Similarly, \cite{kherbache2021digital} constructs a DT model of the network topology for the IIoT to implement a close-loop network management mechanism. It is not surprising that the proposed frameworks use a SDN controller, in a similar way as the SDN in 5G networks, for communicating data between the physical IIoT network and the virtual instance of DT models. This shows that the architecture of DT frameworks is converging to a similar architecture as 5G networks.

\section{Envisioned DT Architecture} \label{sec:arch}
\begin{figure}[t!]
	\centering
		\includegraphics[width=0.5\textwidth]{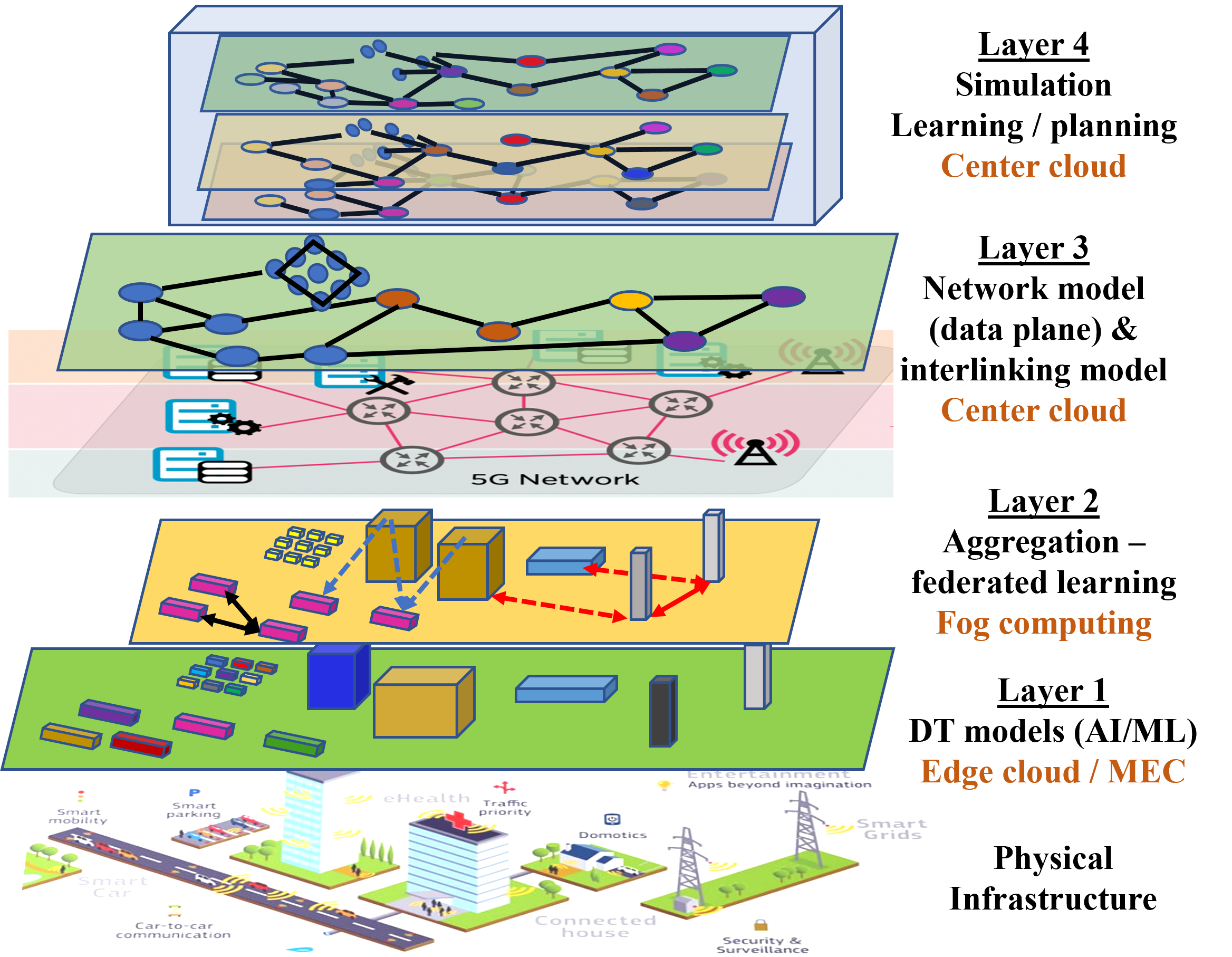}
	\caption{A layered architecture of DT framework with hierarchical deployment on edge, fog and cloud computing. }
	\label{fig:arch}
\end{figure}

The conceptual architecture of our envisioned DT framework is shown in Fig. \ref{fig:arch} that adopts a similar structure as 5G network slicing. To distinguish from 5G, we refer to this architecture as \textit{service slicing}. This terminology is more generic as it refers to control services on a wide range of physical systems as compared with the networking function of 5G. The concept of service in the DT can also include the networking service of 5G networks. In this perspective, a DT framework can model a 5G network for learning the optimal control policy, e.g. for the RIC functions in the O-RAN architecture. 

The deployment of control processes based on DT for 5G intelligent controller (RIC) can serve a twofold purpose. First, the DT framework provides the RIC with simulation environments that can be used for Monte Carlo planning. This learning paradigm is not only beneficial in faster convergence of policy learning but also helps in predictive prevention and event detection. Second, the DT-based controller can be used for realizing off-policy reinforcement learning. The RIC implements the \textit{behavior} policy that controls the (physical) 5G network. The DT deploys the \textit{target} policy that is updated in the virtual space based on the observations from the (physical) network. The off-policy technique can help in converging to the optimal policy rather than a sub-optimal solution.

\begin{figure}[t!]
	\centering
		\includegraphics[width=0.4\textwidth]{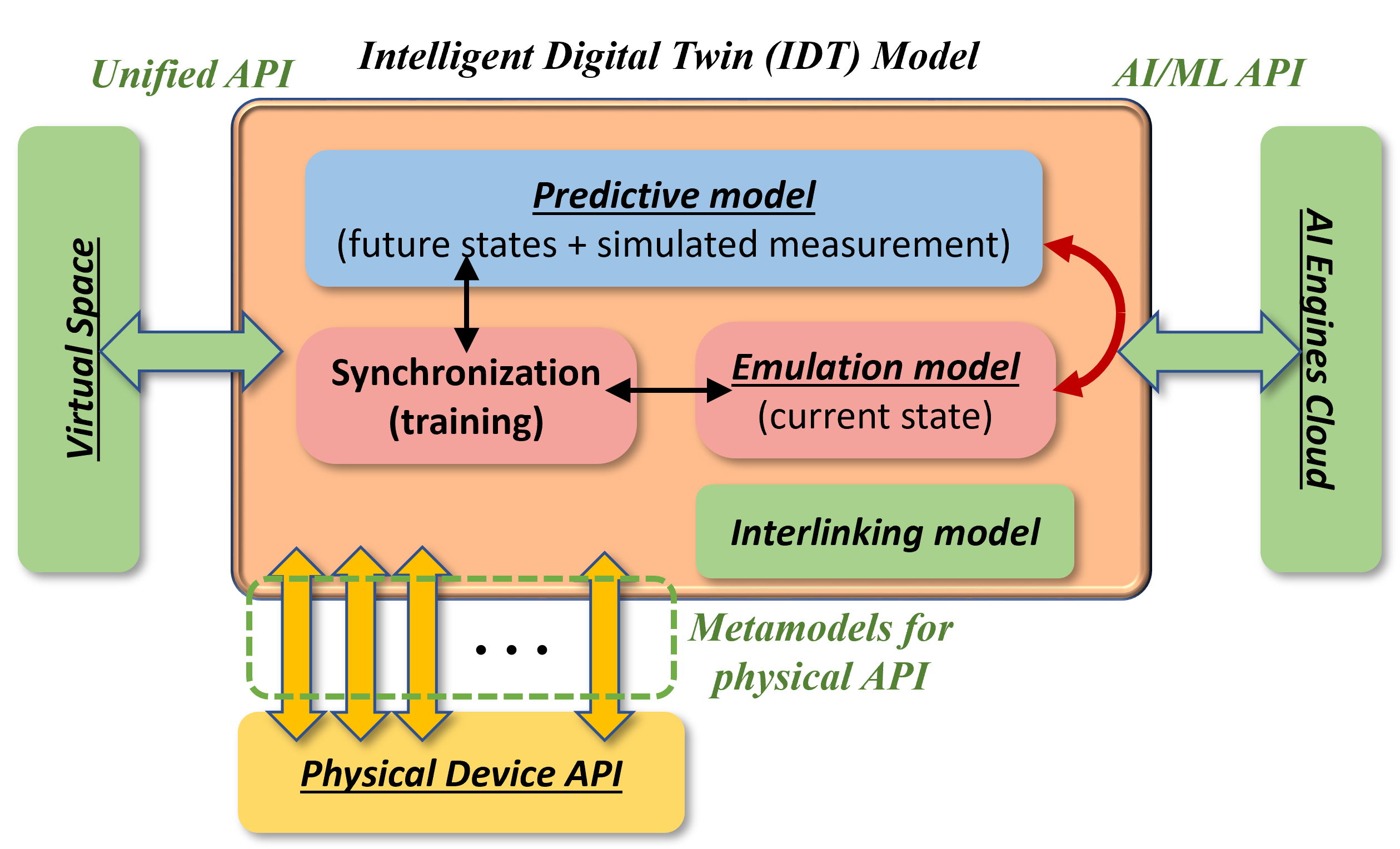}
	\caption{ML-based intelligent DT model for current state (emulation) and environment interactions (predictive model). }
	\label{fig:idt}
	\vspace{-0.4cm}
\end{figure}

The first layer in the proposed DT framework consists of DT models of individual physical entities within their local environment. As shown in Fig. \ref{fig:idt}, this layer directly communicates with the physical space comprising of two sets of physical entities. The first set includes the entities under control which also interact with a physical environment (making measurements from and/or sending actuation to the environment). The second physical set includes the communication infrastructure that can be optionally controlled within the DT framework. 

The DT model of a physical entity, as shown in Fig. \ref{fig:idt}, consists of three main components: emulation model, predictive model, and synchronization engine. The emulation model represents the state and the behavior (interactions) of the entity in a given environment. The state can be a scalar value (e.g., battery percentage of an IoT device) or a multi-dimensional variable (e.g., a vector of geo-spatial state, data rate, security posture, computational and storage capacity). The predictive model represents the dynamics of the local environment interacting with the entity. Given the environment dynamics, the future states and interactions of the physical entity can be predicted, thus the name predictive model. The synchronization engine (or training in the context of AI/Ml models) updates the models using measurements from the environment in real-time.

The second layer of the envisioned DT framework in Fig. \ref{fig:arch} is model aggregation. This layer serves a twofold objective. First, it constructs the DT model of a physical system comprising of smaller devices or models. 
The second objective of this layer is aggregating learning experiences of distributed DTs. Every DT learns a model for the dynamics of its \textit{local} environment. While a DT might observe a limited subset of states of the physical space over a given time. aggregation of states (models) observed by many DTs can provide a wider \textit{visibility} of the space. This is especially important for predictive and simulation capabilities of a DT framework. With learning experience aggregation, a DT model can generate a high-fidelity simulated experience and state prediction even for environments that have not been directly observed by a single DT.

Federated learning is a popular paradigm for aggregating experiences of distributed agents for learning a common model under the constraints of privacy and communication efficiency. The privacy constraint refers to avoiding sharing of local data available only to single agents and preventing exposure of their individual learning (inference) models. Since federated learning does not require sharing raw data at a (logically) centralized training unit, it can result in significant communication efficiency while achieving high accuracy in learning models. 
Model aggregation via federated learning (or any plausible communication-efficient distributed learning technique) is a critical component of the envisioned DT framework. It is a key function for constructing a high-fidelity simulation and prediction models based on limited local observations of individual DTs. Adopting a logically centralized SDN controller for the DT framework facilitates implementation of federated learning.

At the third layer of the DT framework in Fig. \ref{fig:arch}, networking and interlinking models are constructed. As discussed earlier, the physical space comprises of a set of distributed physical devices and processes under control (e.g., smart grid) and a data network infrastructure (e.g., 5G IoT). At this layer of the DT framework, the two physical systems are represented by two virtual models (e.g., interlinking model of smart grid and network model of 5G IoT). 
The interlinking model provides a high-level representation of the controlled system based on local observations of physical entities. Several control processes, corresponding to different service slices, might be associated with different interlinking models. In the example of a smart grid, one service slice monitors energy usage patterns and controls load balancing while a separate service slice implements anomaly detection and cybersecurity controls. The load balancing service uses an interlinking model that describes load profile and power distribution system. The cybersecurity service might employ a model describing grid traffic.

An alternative view of interlinking model is a representation of spatial information about the environment. This is contrast with DT models (associated with individual physical entities) that represent temporal information of the environment state. Graph neural networks (GNN) are promising solutions for constructing interlinking models. The capability of GNN in modeling spatial information has been commonly used in statistical learning techniques for resource allocation in wireless networks \cite{eisen2020optimal}. An important and desirable property of GNNs is the permutation equivariance which makes them capable of exploiting symmetrical properties of the graph and generalize the learned features from local environments to different regions of the physical space. 

The fourth layer of the DT framework implements a simulation environment consisting of several \textit{virtual planes}. Every plane models the physical space at different states. One virtual plane represents the current state of the physical space and can be used for online heuristic search as in learning techniques based on Monte Carlo tree search (MCTS). Additional virtual planes simulate the physical space used in reinforcement learning based on Monte Carlo simulations. Off-policy learning and control engines can also employ dedicated virtual planes for test and verification (cross-validation) before deploying the target policy over the physical space. This is a significant advantage of a DT framework that enables verification of control policies with extensive test plans in real-time without affecting the physical space but with live models mimicking the actual behavior of the space.

The control and decision-making process in AI-enabled DT frameworks rely on live ML models of a physical space for optimal operation. However, emerging adversarial attacks on ML models can lead to concerns about the robustness of a DT framework in adversarial environments. Hence, we envision that emerging DT frameworks would incorporate an assessment service that monitors and evaluates the robustness of underlying DT models against different attacks, including model poisoning, evasion, extraction and inference attacks. We call this service the \textit{adversarial twin} that employs a virtual plane to simulate adversarial attacks on different layers of the DT framework and assess the robustness and security of control processes under different test plans.

The adversarial twin constructs multiple virtual planes in the simulation environment to assess the robustness of DT models and the control process to different attacks discussed above. It can employ adversarial learning techniques to find the weakest link of the system. In the poisoning plane, the adversarial DT learns samples that result in largest deviations of DT models. The evasion plane similarly learns perturbation in measurements that causes false alarms in decision engines. The extraction plane uses extraction models to verify successful simulation of the true states in the physical space. An additional virtual plane simulates the effect of these attacks on the overall control process. These simulations can be used as design guidelines to consolidate the robustness of ML models and monitor adversarial data. 

\section{Research Directions for Security} \label{sec:directions}
A DT framework can be considered as a real-time data-driven modeling of a large-scale distributed physical system. A recent body of research has employed DT to implement model-based event detection processes for CPS. A prominent example is anomaly detection and cybersecurity models in smart grids. An interesting capability of DT in this research area is prediction of future events and failures. This capability can help realize an extension of event/anomaly (intrusion in the context of cybersecurity) detection systems from real-time to the future. 

A DT framework can significantly facilitate system planning and reduce the cost and time of design, development, test and verification. Important applications include integration of renewable energy sources with the smart grid and planning of 5G base-stations for maximizing connectivity in high dynamic environments (e.g., VANETs). For the purpose of planning, DT models representing new physical entities (renewable sources or 5G base-stations) can be inserted to the framework without installing the actual device and equipment in the physical space. The planning DT models can be synchronized with software-based simulators (representing the design) or even physical twins operating in a lab environment. 

The emerging applications of DT can be divided into two broad categories of implementing distributed control and event detection processes over network. The latter has been commonly exploited in cybersecurity applications for intrusion detection and prevention systems (IDS/IPS). Using DT in realizing distributed control is also providential for next-generation cybersecurity systems based on zero trust architecture (ZTA). The problem addressed by a ZTA is implementation of a distributed mechanism for network access control (NAC). The core function of a ZTA is a control process for authorizing accesses to data and network resources, in potentially untrusted network environments. A review of ZTA, its application in beyond 5G networks and the role of machine learning in implementing ZTA is available at \cite{ramezanpour2021intelligent}.

In classical cybersecurity models, authentication (user/device identification) provides a trust basis for granting access to a network resource (based on defined security groups). However, in a ZTA model, a successful authentication does not grant trust to a user, thus the name zero trust. Instead, the NAC requires a model for the security state of the entire network to make decisions on granting or denying individual accesses (from untrusted users). For this purpose, the NAC requires real-time monitoring of all network assets, including users, devices, data, applications, and network environment. A model for security state of assets, using security analytics (data-driven modeling), and also the network interactions (interlinking model), is required for trust evaluation. An automated control process, in the policy decision point (PDP), uses the security analytics and results of event detection engines, to grant or deny access to a network resource. A comparison of these requirements with the capabilities of DT discussed in Section \ref{sec:capabilities} reveals the moment of DT for rapid and reliable adoption of ZTA.

\section{Conclusions} \label{sec:conclusions}
Digital twin (DT) is becoming an inevitable solution for implementation of complex control processes for distributed systems with observations over network. A DT framework paves the path for realization of online learning and planning techniques for exploration and verification of control policies. Its capability in predictive modeling, based on real-time monitoring and data management, allows extensive test and verification of control policies before deployment on the physical space. Hence, it enables deployment of innovative technologies, with lower CAPEX and OPEX, without paying the cost of unprecedented failures. In this position paper, we elucidated unique characteristics of a DT framework that enables realization of these promises. We established a conceptual architecture for a DT framework that facilitates implementation of DT on various systems for realizing critical functions such as event/anomaly detection, data-driven model construction, simulation environments for heuristic search and online Monte Carlo planning.


\bibliographystyle{ACM-Reference-Format}
\bibliography{sample-base}


\end{document}